\documentclass{easychair}

\usepackage{amsmath}
\usepackage{amssymb}
\usepackage{amsthm}
\usepackage{doc}
\usepackage{xspace}
\usepackage{ifthen}

\newcommand\isaforversion{2.33}

\newcommand\isaforbase{http://cl-informatik.uibk.ac.at/isafor/v\isaforversion}
\newcommand\isaforlink[3]{%
  \href{\isaforbase/IWC2018/#1.html\##2}%
  {#3}}

\newcommand{\isafor}{\textsf{Isa\kern-0.2exF\kern-0.2exo\kern-0.2exR}%
\xspace}
\newcommand\ceta{\textsf{C\kern-0.2exe\kern-0.5exT\kern-0.5exA}\xspace}
\newcommand{\maedmax}{\textsf{M{\ae}dMax}\xspace}
\newcommand{\concon}{\textsf{ConCon}\xspace}
\newcommand{\oKB}{\textsf{oKB}\xspace}

\newcommand{\m}[1]{\mathsf{#1}}
\newcommand{\mc}[1]{\mathcal{#1}}
\newcommand{\EE}{\mc{E}}
\newcommand{\FF}{\mc{F}}
\newcommand{\RR}{\mc{R}}
\renewcommand{\SS}{\mc{S}}
\newcommand{\TT}{\mc{T}}
\newcommand{\VV}{\mc{V}}

\newcommand{\Pos}{\mc{P}\m{os}}
\newcommand{\PosF}{\Pos_\FF}
\newcommand{\ECP}[1][\succ]{\textup{CP}_{#1}}
\newcommand{\join}{\downarrow}
\newcommand{\conv}{\leftrightarrow^*}
\newcommand{\from}{\leftarrow}
\newcommand{\kbo}{\m{kbo}}
\newcommand{\lpo}{\m{lpo}}
\newcommand\REgt{\RR \cup \EE^>}

\newcommand\symcl[1]{#1^{\leftrightarrow}}
\colorlet{fun}{black!65!green}
\colorlet{var}{black!30!orange}
\newcommand\fs[2][]{\mathsf{\textcolor{fun}{#2_{#1}}}}
\newcommand\vs[2][]{\mathit{\textcolor{var}{#2_{#1}}}}
\newcommand{\0}{\fs{0}}
\newcommand\tru{\fs{true}}
\newcommand\X{\vs{x}}
\newcommand\Y{\vs{y}}
\newcommand\U{\vs{u}}
\newcommand\V{\vs{v}}
\newcommand\Q{\vs{q}}
\newcommand\R{\vs{r}}

\newcommand\fsdivname{\fs{\div}}
\newcommand\fsgtname{\fs{>}}
\newcommand\fslename{\fs{\leq}}
\newcommand\fsminusname{\fs{-}}
\newcommand\fsdiv[2]{#1 \mathbin{\fsdivname} #2}
\newcommand\fsgt[2]{#1 \mathbin{\fsgtname} #2}
\newcommand\fsle[2]{#1 \mathbin{\fslename} #2}
\newcommand\fsminus[2]{#1 \mathbin{\fsminusname} #2}

\newcommand\eq{\approx}
\newcommand{\pair}[2]{\langle#1,#1\rangle}

\newcommand\EMPH[1]{\textbf{#1}}

\newcommand\cops[1]{\href{http://cops.uibk.ac.at/?q=#1}{\texttt{\##1}}}

\newcommand{\doi}[1]{
  \href{http://dx.doi.org/#1}{\nolinkurl{doi:#1}}}

\newcommand\defref[1]{Definition~\ref{def:#1}}
\newcommand\lemref[2][]{%
  Lemma~\ifthenelse{\equal{#1}{}}%
    {\ref{lem:#2}}%
    {\ref{lem:#2}(\ref{lem:#2:#1})}}

\newcommand\secref[1]{Section~\ref{sec:#1}}

\theoremstyle{definition}
\newtheorem{definition}{Definition}

\theoremstyle{theorem}
\newtheorem{lemma}{Lemma}
\newtheorem{theorem}{Theorem}

\title{Certified Ordered Completion%
\thanks{This work is supported by the Austrian Science Fund (FWF): projects T789 and P27502.}}

\author{
Christian Sternagel
\and Sarah Winkler
}

\institute{
  Department of Computer Science \\
  University of Innsbruck, Innsbruck, Austria\\
  \email{\{christian.sternagel|sarah.winkler\}%
  @uibk.ac.at}}
  
\authorrunning{Sternagel and Winkler}

\titlerunning{Certified Ordered Completion}

\begin{document}

\maketitle

\begin{abstract}
On the one hand, ordered completion is a fundamental technique in equational
theorem proving that is employed by automated tools. On the other hand, their
complexity  makes such tools inherently error prone.
As a remedy to this situation we give an Isabelle/HOL formalization of
ordered rewriting and completion that comes with a formally verified
certifier for ordered completion proofs.
By validating generated proof certificates, our certifier increases the
reliability of ordered completion tools.
\end{abstract}

\section{Introduction}
\label{sec:introduction}

Completion has evolved as a fundamental technique in automated reasoning
since the ground-breaking work by Knuth and Bendix~\cite{KB70}.
Its goal is to transform a given set of equations into a terminating and
confluent term rewrite system that induces the same equational theory and can
thus be used to decide equivalence with respect to the 
initial set of
equations. Since the
original procedure can fail if unorientable equations are
encountered, ordered completion was developed to remedy this
shortcoming~\cite{BDP89}. The systems generated by ordered completion
tools are in general only ground
confluent, but this turns out to be sufficient for
practical applications like refutational theorem proving.

Consider for example the following equational system $\EE_0$
which the tool~\maedmax~\cite{WM18}
\begin{xalignat*}{3}
\fsdiv{\X}{\Y} &\eq \pair{\0}{\Y} &
\fsdiv{\X}{\Y} &\eq \pair{\fs{s}(\Q)}{\R} &
\fsminus{\X}{\0} &\eq \X \\
\fsminus{\0}{\Y} &\eq \0 &
\fsminus{\fs{s}(\X)}{\fs{s}(\Y)} &\eq \fsminus{\X}{\Y} &
\fsgt{\fs{s}(\X) }{\fs{s}(\Y}) &\eq \fsgt{\X }{\Y} \\
\fsgt{\fs{s}(\X) }{\0} &\eq \tru &
\fsle{\fs{s}(\X)}{\fs{s}(\Y}) &\eq \fsle{\X}{\Y} &
\fsle{\0}{\X} &\eq \tru
\end{xalignat*}
transforms by ordered completion into the following rules $\RR$ ($\to$) and
equations $\EE$ ($\eq$):
\begin{xalignat*}{4}
\fsminus{\X}{\0}	&\to	\X &
\fsminus{\0}{\X}	&\to	\0 &
\fsminus{\fs{s}(\X)}{\fs{s}(\Y)}	&\to	\fsminus{\X}{\Y} &
\fsdiv{\X}{\Y}	&\to	\pair{\0}{\Y}\\
\fsle{\m{\0}}{\X}	&\to	\tru &
\fsle{\fs{s}(\X)}{\fs{s}(\Y})	&\to	\fsle{\X}{\Y}  &
\fsgt{\fs{s}(\X) }{\0}	&\to	\tru \\
\fsgt{\fs{s}(\X) }{\fs{s}(\Y})	&\to	\fsgt{\X }{\Y} &
\pair{\fs{s}(\X)}{\Y}	&\eq	\pair{\fs{s}(\Q)}{\R} &
\pair{\fs{s}(\Q)}{\R}	&\eq	\pair{\0}{\Y} &
\pair{\0}{\X}	&\eq	\pair{\0}{\Y}
\end{xalignat*}
This system can be used to decide a given ground equation
by checking whether the terms' unique normal
forms (with respect to ordered rewriting) are equal.

Such ground complete systems are useful for other tools, like
\concon~\cite{SM14}---a tool for automatically proving confluence of conditional
term rewrite systems---which employs ordered completion for proving
infeasibility of conditional critical pairs.
In fact, $\EE_0$ from our initial example is the equational system that 
\concon derives from Cops~\cops{361} for that purpose.
The latter models division with remainder,
though the transformation performed by \concon creates some equations
which do not fit into this semantics but are required to decide confluence.

However, automated tools like \concon and \maedmax are complex and
highly optimized.
The produced proofs often comprise hundreds of equations and thousands of steps.
Hence care should be taken to trust the output of such tools.

To 
improve this situation we follow a two-staged certification
approach and first
(1) add the relevant concepts and results to a formal library, and then
(2) use code generation to obtain a trusted certifier.
More specifically, our contributions are as follows:
\begin{itemize}
\item
Regarding stage (1), we extended the \textsl{\EMPH{Isa}belle
\EMPH{F}ormalization \EMPH{o}f
\EMPH{R}ewriting}\footnote{\url{http://cl-informatik.uibk.ac.at/isafor}}
(\isafor) by ordered rewriting and a generalization of the ordered completion
calculus \oKB~\cite{BDP89}, and proved the latter correct for finite runs using
ground-total reduction orders (\secref{okb}).  Moreover, we established
ground-totality of the lexicographic path order and the Knuth-Bendix order.

\item
With respect to stage (2),
we extended the XML-based \emph{certification problem format} (CPF for short)
\cite{ST14}
by certificates comprising the initial
equations, the resulting system along with a reduction order, and a stepwise
derivation of the latter from the former.
We then formalized check functions that verify that the supplied derivation
corresponds to a valid \oKB run 
whose final state matches the resulting
system (\secref{proof checking}).
As a result \ceta (the certifier accompanying \isafor) can now certify ordered
completion proofs produced by the tool \maedmax~\cite{WM18}.
\end{itemize}

\section{Preliminaries}
\label{sec:preliminaries}

In the sequel we use standard notation from term rewriting~\cite{BN98}.
We consider the \emph{set of all terms} $\TT(\FF,\VV)$ over a signature $\FF$
and an infinite set of variables $\VV$, while $\TT(\FF)$ denotes the
\emph{set of all ground terms}. 
A \emph{substitution} $\sigma$ is a mapping from variables to terms.
As usual, we write $t\sigma$ for the \emph{application} of $\sigma$ to a term $t$.
A \emph{variable permutation} (or \emph{renaming})~$\pi$ is a bijective substitution
such that $\pi(x) \in \VV$ for all $x\in\VV$.
For an equational system (ES) $\EE$ we write $\symcl{\EE}$ to
denote its symmetric closure~$\EE \cup  \{t \eq s \mid s\eq t \in \EE\}$.
For a reduction order $>$ and an ES~$\EE$, the term rewrite system (TRS)~$\EE^>$
consists of all rules~$s\sigma \to t\sigma$ such that $s \eq t \in \EE$ and
$s\sigma > t\sigma$.

Given a reduction order $>$, an \emph{extended overlap} is given by two
variable-disjoint variants~\mbox{$\ell_1 \eq r_1$} and~$\ell_2 \eq r_2$ of
equations in $\symcl{\EE}$ such that $p \in \PosF(\ell_2)$ and $\ell_1$ and
$\ell_2|_p$ are unifiable with most general unifier~$\mu$.
An extended overlap which in addition satisfies $r_1\mu \not > \ell_1\mu$
and $r_2\mu \not > \ell_2\mu$ gives rise to the \emph{extended critical pair}
$\ell_2[r_1]_p\mu \eq r_2\mu$.  The set $\ECP[>](\EE)$ consists of all extended
critical pairs among equations in 
$\EE$.
A TRS~$\RR$ is \emph{(ground) complete} if it is terminating and confluent
(on ground terms).
Finally, we say that a TRS~$\RR$ is a presentation of an ES~$\EE$,
whenever ${\conv_\EE} = {\conv_\RR}$.

\section{Formalizing Ordered Completion}
\label{sec:okb}

We consider the following definition of ordered
completion.

\begin{definition}[Ordered Completion]
\label{def:oKB}
\isaforlink{Ordered_Completion}{ind:oKB'}{%
The inference system \textsf{oKB} of ordered completion
operates on pairs $(\EE,\RR)$ of equations~$\EE$ and rules~$\RR$
over a common signature $\FF$. It consists of the
following inference rules, where $\SS$ abbreviates $\REgt$
and $\pi$ is a renaming.
\begin{center}
\begin{tabular}{@{}lc@{~~}l@{\qquad}lc@{~~}l@{}}
\textsf{deduce} &
$\displaystyle \frac
{\EE,\RR}
{\EE \cup \{ s\pi \eq t\pi \},\RR}$
& if
$s \xleftarrow[\RR \cup \EE]{} \cdot \xrightarrow[\RR \cup \EE]{} t$
&
\textsf{compose} &
$\displaystyle \frac
{\EE,\RR \uplus \{ s \to t \}}
{\EE,\RR \cup \{ s\pi \to u\pi \}}$
& if $t \xrightarrow{}_{\SS} u$
\\ & \\
&
$\displaystyle \frac
{\EE \uplus \{ s \eq t \},\RR}
{\EE,\RR \cup \{ s\pi \to t\pi \}}$
& if $s > t$
&
&
$\displaystyle \frac
{\EE \uplus \{ s \eq t \},\RR}
{\EE \cup \{ u\pi \eq t\pi \},\RR}$
&
if $s \to_{\SS} u$
\\[-.5ex]
\textsf{orient} & & &
\textsf{simplify}
\\[-.5ex]
&
$\displaystyle \frac
{\EE \uplus \{ s \eq t \},\RR}
{\EE,\RR \cup \{ t\pi \to s\pi \}}$
& if $t > s$
&
&
$\displaystyle \frac
{\EE \uplus \{ s \eq t \},\RR}
{\EE \cup \{ s\pi \eq u\pi \},\RR}$
&
if $t \to_{\SS} u$
\\ & \\
\textsf{delete} &
$\displaystyle \frac
{\EE \uplus \{ s \eq s \},\RR}
{\EE,\RR}$ &
&
\textsf{collapse} &
$\displaystyle \frac
{\EE,\RR \uplus \{ t \to s \}}
{\EE \cup \{ u\pi \eq s\pi \},\RR}$
&
if $t \to_{\SS} u$
\end{tabular}
\end{center}
\mbox{}}
\end{definition}
We write $(\EE,\RR) \vdash (\EE',\RR')$ if $(\EE',\RR')$ is obtained
from $(\EE,\RR)$ by employing one of the above inference rules.
A finite sequence of inferences
$
(\EE_0,\varnothing) \vdash (\EE_1,\RR_1) \vdash \cdots \vdash (\EE_n,\RR_n)
$
is called a \emph{run}. \defref{oKB} differs from the original
formulation of ordered completion~\cite{BDP89} in two ways.
First, \textsf{collapse} and \textsf{simplify} do not require an encompassment
condition. This omission is possible since we only consider
\emph{finite} runs.
Second, we allow variants of rules and equations to be added.
This relaxation tremendously simplifies certificate generation in tools,
where facts are renamed upon generation
to avoid the maintenance and processing of many renamed versions of one
equation.

The following inclusions express
straightforward properties of \oKB.

\begin{lemma}
\label{lem:oKB less}
\isaforlink{Ordered_Completion_Impl}{lem:oKB'_rtrancl_less}{%
If $(\EE,\RR) \vdash^* (\EE',\RR')$ then $\RR \subseteq {>}$
implies $\RR' \subseteq {>}$.
}
\qed
\end{lemma}

\begin{lemma}
\label{lem:oKB conversion}
\isaforlink{Ordered_Completion_Impl}{lem:oKB_steps_conversion_permuted}{%
If $(\EE,\RR) \vdash^* (\EE',\RR')$ then the conversion equivalence
${\conv_{\EE\cup\RR}} = {\conv_{\EE'\cup\RR'}}$ holds.
}
\qed
\end{lemma}

The following abstract result is the key ingredient to our proof of
ground completeness.

\begin{lemma}
\label{lem:GCR ordstep}
\isaforlink{Ordered_Rewriting}{lem:GCR_ordstep}{%
Let $\EE$ be an ES and $>$ a reduction order such that 
$s > t$ or $t \eq s \in \EE$ holds for all
$s \eq t \in \EE$.
If for all $s \eq t \in \ECP[>](\EE)$ we have $s \join_{\EE^>} t$ 
or there is some $s' \eq t' \in \symcl{\EE}$ such that
$s \eq t = (s' \eq t')\sigma$ then $\EE^>$ is ground complete.
}
\qed
\end{lemma}

In combination, Lemmas~\ref{lem:oKB less}, \ref{lem:oKB conversion}, and
\ref{lem:GCR ordstep} allow us to obtain our main correctness result:
acceptance of a certificate by our check function 
implies
that $\REgt$
is a ground complete presentation of $\EE_0$. For simplicity's sake, we give
only the corresponding high-level result (that is, not mentioning our concrete
implementation):
\begin{theorem}
\label{thm:correctness}
\isaforlink{Check_Completion_Proof}{lem:check_ordered_completion_proof_sound}{%
Suppose 
$(\EE_0,\varnothing) \vdash^* (\EE,\RR)$ was obtained
using a ground-total reduction order $>$
with minimal constant $c$ and  for all
$s \eq t \in \ECP[>](\symcl{\EE} \cup \RR)$ either
$s \join_{\REgt} t$, or 
$s \eq t = (s' \eq t')\sigma$ for some $s' \eq t' \in \symcl{\EE}$.
Then ${\conv_{\EE_0}} = {\conv_{\RR\cup\EE}}$ and $\REgt$ is ground complete.}\qed
\end{theorem}
This result employs the following sufficient condition for ground
completeness:
all critical pairs are joinable or instances of equations already present.
In fact, this is not a necessary condition.
%
Martin and Nipkow~\cite{MN90} gave examples of ground confluent systems that 
do not satisfy this condition, and presented a stronger criterion.
However, ground confluence is known to be undecidable even for terminating
TRSs~\cite{KNO90}, hence no complete criterion can be implemented.

\paragraph{Ground-total reduction orders.}
Ground confluence crucially relies on ground-total reduction orders. Our \isafor
proofs of the following results follow the standard textbook
approach~\cite{BN98}.

\begin{lemma}
\label{lem:LPO gtotal}
\isaforlink{RPO}{lem:lpo_ground_total}{%
If $>$ is a total precedence on $\FF$ then $>_\lpo$ is total on $\TT(\FF)$.
}\qed
\end{lemma}

\begin{lemma}
\label{lem:KBO gtotal}
\isaforlink{KBO}{lem:S_ground_total}{%
If $>$ is a total precedence on $\FF$ then $>_\kbo$ is total on $\TT(\FF)$.
}\qed
\end{lemma}

In addition, we proved that for any given KBO $>_\kbo$ (LPO $>_\lpo$) defined
over a total precedence $>$ there exists a minimal constant $c$ such that
$t \geqslant_\kbo c$ ($t \geqslant_\lpo c$) holds for all $t \in \TT(\FF)$.

\section{Checking Ordered Completion Proofs}
\label{sec:proof checking}

While \ceta 
has supported
certification of standard completion for quite some
time~\cite{ST13},
certification of ordered completion proofs is considerably more intricate.
For standard completion, the certificate contains the initial set of equations
$\EE_0$, the resulting TRS~$\RR$ together with a termination proof, and stepwise
$\EE_0$-conversions from $\ell$ to $r$ for each rule $\ell \to r \in \RR$. The
certifier first checks the termination proof to guarantee
termination of $\RR$. This allows us to establish confluence of $\RR$ by ensuring
that all critical peaks are joinable. At this point it is easy to verify
${\conv_{\EE_0}} \subseteq {\conv_\RR}$: for each equation $s \eq t \in \EE_0$
compute the $\RR$-normal forms of $s$ and $t$ and check for syntactic equality.
The converse inclusion
${\conv_\RR} \subseteq {\conv_{\EE_0}}$
is taken care of by the provided $\EE_0$-conversions.
Overall, we obtain that $\RR$ is a complete presentation of $\EE_0$ without
mentioning a specific inference system for completion.

Unfortunately, the same approach does not work for ordered completion:
The inclusion ${\conv_{\EE_0}} \subseteq {\conv_{\RR \cup \EE}}$ cannot be
established by rewriting equations in $\EE_0$ to normal form, since they
may contain variables but $\REgt$ is only ground confluent.
Therefore, we instead ask for certificates that contain the input equalities
$\EE_0$, the resulting equations and rules $(\EE,\RR)$, the reduction
order $>$, and a sequence of inference steps according to \defref{oKB}.
A valid certificate ensures (by \lemref{oKB conversion}) that the
relations $\conv_{\EE_0}$ and $\conv_{\RR \cup \EE}$ coincide.
\smallskip

\newcommand{\orientlr}{\textsf{orient}$_{\m{lr}}$\xspace}%
\newcommand{\orientrl}{\textsf{orient}$_\m{rl}$\xspace}%
The certificate corresponding to our initial example contains the equations
$\EE_0$, the resulting system $(\EE,\RR)$, and the reduction order
$>_\kbo$ with precedence $\fsgtname > \fs{s} > \fslename > \tru > \fsminusname > \fsdivname > \fs{p} > \0$,
$w_0 = 1$, and 
$w(\0) = 2$, $w(\fsdivname) = w(\tru) = w(\fs{s}) = 1$,
and all other symbols having weight 0.
In addition, a sequence of inference steps explains how $(\EE,\RR)$ is obtained from
$\EE_0$:
\smallskip

\noindent
\parbox{2cm}{\textsf{simplify$_{\m{left}}$}} 
 $\fsdiv{\X}{\Y} \eq \pair{\fs{s}(\Q)}{\R}$ to $\pair\0\Y \eq
 \pair{\fs{s}(\Q)}{\R}$\\
\parbox{2cm}{\textsf{deduce}}
 $\pair\0\X \from \pair{\fs{s}(\U)}{\V} \to \pair{\0}{\Y}$ \\
\parbox{2cm}{\textsf{deduce}}
 $\pair{\fs{s}(\X)}{\Y} \from \pair\0\U \to \pair{\fs{s}(\Q)}{\R}$\\
\parbox{2cm}{\textsf{deduce}}
 $\fsgt{\X}{\Y} \from \fsgt{\fs{s}(\X)}{\fs{s}(\Y}) \to
 \fsgt{\fs{s}(\fs{s}(\X))}{\fs{s}(\fs{s}(\Y}))$\\
\parbox{2cm}{\textsf{deduce}}
 $\fsgt{\fs{s}(\fs{s}(\X))}{\fs{s}(\0}) \from \fsgt{\fs{s}(\X)}{\0} \to \tru$ \\
\parbox{2cm}{\orientrl}
 $\fsle{\0}{\X} \to \tru$ \\
\parbox{2cm}{\orientlr}
 $\fsgt{\fs{s}(\fs{s}(\X))}{\fs{s}(\0}) \to \tru$\\
\parbox{2cm}{\orientrl}
 $\fsgt{\fs{s}(\X)}{\fs{s}(\Y}) \to \fsgt{\X}{\Y}$\\
\parbox{2cm}{\orientlr}
 $\fsgt{\fs{s}(\X)}{\0} \to \tru$ \\
\parbox{2cm}{\orientrl}
 $\fsgt{\fs{s}(\fs{s}(\X))}{\fs{s}(\fs{s}(\Y})) \to \fsgt{\X}{\Y}$ \\
\parbox{2cm}{\orientrl}  $\fsminus{\X}{\0} \to \X$ \\
\parbox{2cm}{\orientlr}  $\fsdiv{\X}{\Y} \to \pair{\0}{\Y}$\\
\parbox{2cm}{\orientrl}  $\fsminus{\fs{s}(\X)}{\fs{s}(\Y)} \to \fsminus{\X}{\Y}$\\
\parbox{2cm}{\orientrl}  $\fsminus{\0}{\X} \to \0$ \\
\parbox{2cm}{\orientrl}  $\fsle{\fs{s}(\X)}{\fs{s}(\Y}) \to \fsle{\X}{\Y}$ \\
\parbox{2cm}{\textsf{collapse}}
 $\fsgt{\fs{s}(\fs{s}(\X))}{\fs{s}(\fs{s}(\Y})) \to \fsgt{\X}{\Y}$ to
 $\fsgt{\fs{s}(\X)}{\fs{s}(\Y}) \eq \fsgt{\X}{\Y}$\\
\parbox{2cm}{\textsf{simplify$_{\m{left}}$}}
 $\fsgt{\fs{s}(\X)}{\fs{s}(\Y}) \eq \fsgt{\X}{\Y}$ to $\fsgt{\X}{\Y} \eq
 \fsgt{\X}{\Y}$ \\
\parbox{2cm}{\textsf{collapse}}
 $\fsgt{\fs{s}(\fs{s}(\X))}{\fs{s}(\0}) \to \tru$ to $\fsgt{\fs{s}(\X)}{\0}
 \eq \tru$ \\
\parbox{2cm}{\textsf{simplify$_{\m{left}}$}}
 $\fsgt{\fs{s}(\X)}{\0} \eq \tru$ to $\tru \eq \tru$\\
\parbox{2cm}{\textsf{delete}}  $\fsgt{\X}{\Y} \eq \fsgt{\X}{\Y}$\\
\parbox{2cm}{\textsf{delete}}  $\tru \eq \tru$

Given such a certificate, \ceta checks that the provided sequence of
inferences forms a run $(\EE_0\pi,\varnothing) \vdash^* (\EE,\RR)$
for some renaming $\pi$.
Verifying the validity of individual inferences involves checking side conditions
such as orientability of a term pair in an orient step with respect to the given
reduction order.
Then it is checked that $\REgt$ is ground confluent according to the
criterion of Theorem~\ref{thm:correctness}.
Finally, it is  ensured that the given reduction order $>$ has a total precedence
(and is admissible, in the case of KBO). As usual in \ceta, error messages are printed
if one of these checks fails, pointing out the reason for the proof being
rejected.

\section{Conclusion}
\label{sec:conclusion}
We presented our formalization of ordered completion in \isafor, which enables
\ceta (starting with version~\isaforversion) to certify ordered completion
proofs. To the best of our knowledge, \ceta thus constitutes the first formally
verified certifier for ordered completion. 

Together with Hirokawa and Middeldorp we reported on another Isabelle/HOL
formalization of ordered completion \cite{HMSW17}.  The main difference to our
current work is that this other formalization is based on a more restrictive
inference system of ordered completion that also covers infinite runs, while
we restrict to finite runs in the interest of certification.
Indeed every finite run akin to \cite[Definition 18]{HMSW17} is also a run 
according to Definition~\ref{def:oKB}, while the inference sequence in our 
running example is not possible in the former setting.

As future work, we plan to add more powerful criteria for ground confluence to \isafor,
and support equational disproofs based on ground complete systems in \ceta.
To that end, it would be useful to also support narrowing in \ceta.
Certified equational disproofs could in turn be used to certify confluence 
proofs by \concon which rely on infeasibility of conditional critical pairs.

\bibliographystyle{abbrv}
\bibliography{references}

\end{document}